\documentstyle[12pt]{article}
\def\b{\bar}
\def\d{\partial}
\def\D{\Delta}
\def\cD{{\cal D}}

\def\l{\lambda}

\def\m{\mu}
\def\n{\nu}

\def\q{\b q}

\def\t{\tau}

\def\~{\tilde}
\def\h{\eta}

\def\bY3{\bar Y_{,3}}
\def\Y3{Y_{,3}}
\def\z{\zeta}
\def\Z{{\b\zeta}}
\def\Y{{\bar Y}}
\def\cZ{{\bar Z}}
\def\`{\dot}
\def\be{\begin{equation}}
\def\ee{\end{equation}}
\def\bea{\begin{eqnarray}}
\def\eea{\end{eqnarray}}

\def\fn{\footnote}

\def\mn{{\mu\nu}}

\begin{document}
\title{Complex Structure of Kerr Geometry and  Rotating
``Photon Rocket'' Solutions.}

\author{Alexander Burinskii\\
Gravity Research Group, NSI Russian
Academy of Sciences,\\
B. Tulskaya 52, 113191 Moscow, Russia}

\maketitle

\begin{abstract}
In the frame of the Kerr-Schild approach, we obtain a generalization of
the Kerr solution to a nonstationary case corresponding to a rotating
source moving with arbitrary acceleration.
Similar to the Kerr solution, the solutions obtained have the geodesic and
shear free principal null congruence. The current parameters of the
solutions are determined by a complex retarded-time construction via a given
complex worldline of source.  The real part of the complex worldline defines
the values of the boost and acceleration while the imaginary part controls
the rotation.  The acceleration of the source is accompanied by a lightlike
radiation along the principal null congruence.  The solutions obtained
generalize to the rotating case the known Kinnersley class of the "photon
rocket" solutions.  \end{abstract}

PACS number(s): 04.20.Jb 04.20.-q 97.60.Lf 11.27.+d

\section{Introduction}
The nonstationary gravitational fields from moving sources can be described
by a retarded-time scheme which is a generalization of the Lienard-Wiechard
construction of classical electrodynamics. In particular,
the known Kinnersley "photon rocket" solutions \cite{Kin,KraSte}
represent the nonstationary generalizations of the Schwarzschild
 black hole.  The parameters of the Kinnersley solution are
determined by a given worldline of a moving source and are expressed via the
retarded-time position, boost and acceleration of the source.
Due to acceleration,
the Kinnersley solutions are accompanied by a lightlike radiation.

In this work we generalize this solution to the rotating case corresponding
to the twisting Kerr geometry.

Our treatment is based on the  Newman complex representation of
the Kerr geometry \cite{LinNew}, which is generated by a complex world
line of source $x_0(\t)$ moving in $CM^4$. This representation was used
 to construct the retarded-time scheme \cite{LinNew}
and to obtain solutions of the linearized Einstein equations \cite{KN}.

In the present work we construct such a retarded-time scheme in the
Kerr-Schild formalism \cite{DKS}. It allows us to use the effective Kerr
theorem \cite{KraSte,Pen} and obtain  exact nonstationary
 solutions having the twisting, geodesic and shear free Kerr congruence.
These solutions represent a natural generalization of the Kinnersley class.

Complex structure of Kerr geometry plays a central role in
our treatment.
We obtain a link between the Kerr theorem and the complex worldline of the
Kerr
 source and determine  the Kerr congruence as a complex retarded-time field.
 Real solutions are formed as a real slice of the complex structure and
  determined by a real retarded-time parameter.

\section{ Complex structure of Kerr geometry and
retarded-time construction.}

Electrodynamic analogue of the Kerr solution was obtained
by Appel in 1887 (!) \cite{App}.
A point-like charge $e$, placed on the complex Z-axis $(x_0,y_0,z_0)=
(0,0, ia)$ gives the Appel potential
$\phi_a = Re \  e/{\tilde r},$
where $\tilde r$ is the Kerr complex radial coordinate
$\tilde r= PZ^{-1}= r+ i a \cos\theta$,
and $r$, $\theta$ are the oblate spheroidal coordinates.
It may be expressed in the usual Cartesian coordinates $x,y,z,t$  as
$\tilde r = [(x-x_0)^2 + (y-y_0)^2 + (z-z_0)^2]^{1/2}
= [x^2 + y^2 + (z-ia)^2]^{1/2}, $
that corresponds to a shift of the source in complex direction
$(x_o, y_o , z_o ) \rightarrow (0,0,ia)$, and
can be considered as a mysterious "particle" propagating
along a {\it complex} world-line $x_0^\mu (\tau)$ in $CM^4$.

The Kerr-Newman solution has just the same origin
and can be described by means of a complex retarded-time construction
as a field generated by such a complex source
\cite{LinNew,IvBur,Bur1,BKP,BurMag}.

Similar to the usual retarded-time scheme, a {\it complex } retarded time
$\tau =t+i\sigma$
is determined by the family of complex light cones emanating from the points
of the complex worldline $x_0^\mu (\tau)$.

The complex light cone with the vertex at
some point $x_0$ of the
complex worldline $x_0^\mu(\tau) \in
CM^4$: $\quad (x_\mu - x_{0 \mu})(x^\mu -x_0^\mu) = 0 ,$
can be split into two families of  null
planes:
"left" planes
\be
 x_L = x_0(\tau) + \alpha e^1 + \beta e^3 \label{L}
\ee
spanned
by null vectors $e^1 (Y)$ and $e^3 (Y, \tilde Y)$,
and"right"planes
\be
 x_R =
x_0(\tau) + \alpha e^2 + \beta e^3, \label{R}
\ee
spanned by null vectors $e^2$ and $e^3$.

 The Kerr congruence $\cal K$ arises as a real slice of the family
of the "left" null planes ($Y=const.$) of the complex light cones whose
vertices lie on the complex worldline $x_0(\tau)$.

\section{Kerr-Schild formalism and the Kerr theorem}

In the Kerr-Schild backgrounds congruence $\cal K$ determines the ansatz
\be
g_{\m\n} =
\h_{\m\n} + 2 h e^3_{\m} e^3_{\n}, \label{(1.5)}
\ee
where $ \h_\mn $ is auxiliary Minkowski metric,
and $e^3$ is principal null direction (PNC) of Kerr geometry.
The PNC is null with respect to the auxiliary Minkowski spacetime as well as
to the metric $g_\mn$.  \fn{ We follow the notations of work \cite{DKS} and
use signature $(-+++)$.  The Kerr-Schild null tetrad is completed as
follows:
 $e^1 = d \zeta - Y dv, \quad  e^2 = d \bar\zeta -  \bar Y dv, \quad e^4 =dv
+
h e^3.$ The dual tetrad $e_a$ is determined by permutations:
$e^{1}\to e_{2}, e^{2}\to e_{1}, e^{3}\to e_{4}, e^{4}\to e_{3}$. The
tetrad derivatives are defined by $,_a \equiv \d _a = e^\m _a \d_\m$.}

The treatment via complex worldline can be related to the Kerr theorem
by setting up a correspondence between parameters of the worldline and
parameters of a generating function $F$ of the Kerr theorem.

Traditional formulation of the Kerr theorem is as follows.

Any  geodesic and shear-free null congruence in
Minkowski space is defined by a function $Y(x)$ which is a solution of the
equation
\be         F  = 0 ,                            \label{(1.1)}\ee
where   $F (\l_1,\l_2,Y)$   is an arbitrary analytic
function of the projective twistor coordinates $
Y,\qquad \l_1 = \z - Y v, \qquad \l_2 =u + Y \Z .$

The Kerr congruence $\cal K$ is defined then by the vector field
\be
 e^3 = du+ \Y d \z  + Y d \Z - Y \Y d v \ ,
\label{(1.8)}
\ee
where
$2^{1\over2}\z = x+iy ,\quad 2^{1\over2} \Z = x-iy , \quad
2^{1\over2}u = z + t ,\quad 2^{1\over2}v = z - t  $
are the null Cartesian coordinates.

In the Kerr-Schild approach the Kerr theorem acquires  more broad contents
\cite{DKS,IvBur,BKP}.  It allows one to obtain the position of singular
lines,
caustics of the PNC, as a solution of the system of equations
$ F=0;\quad d F /d Y =0, $
and to define some important parameters of the solution:
\be \tilde r = - \quad d F / d Y , \qquad P = \d_{\l_1} F - \Y \d_{\l_2} F.
\label{PF} \ee The parameter $\tilde r$ characterizes a complex radial
 distance, and for the Kerr solution it is a typical complex combination
$\tilde r= r+ia \cos\theta$.  Parameter $P$ is connected with the boost of
the source.

Stationary congruences, having Kerr-like singularities contained in
a bounded region, have been considered in papers \cite{IvBur,KerWil,BurMag}.
It was shown that in this case function $F$ must be at most quadratic in
$Y$,
\be
F \equiv a_0 +a_1 Y + a_2 Y^2 + (q Y + c) \l_1 - (p Y + \q) \l_2,
\label{FK}
\ee
where coefficients $ c$ and $ p$ are real constants
and $a_0, a_1, a_2,  q, \q, $  are complex constants.
Writing the function F in the form
$F = A  Y^2 + B Y + C, $
one can find solutions of the equation $F=0$ for the
function $Y(x)$
\be
 Y_{1,2} = (- B \pm \D )/2A, \label{Y12}
\ee
where $ \D = (B^2 - 4AC)^{1/2}.$
These two roots define two PNCs of the Kerr geometry.
From  (\ref{PF}) one has
\be
\tilde r = - \d F /\d Y= -2AY -B = \mp \D,
\label{tr1}
\ee
and $P=pY\Y  + \q \Y + qY +c . $

On the other hand, the stationary and boosted Kerr geometries are described
by a  straight complex worldline with a real 3-velocity $\vec v$ in $CM^4$:
\begin{equation}
x_0^\m (\t) = x_0^\m (0) + \xi^\m \t; \qquad \xi^\m = (1,\vec v)\ .
\label{dec}
\end{equation}
It was shown in \cite{BurMag} that parameters
$p, c, q, \q$ are related to parameters of complex worldline $\d _\t
x_0(\t)=\xi ^\mu$, or to the boost of the source.  Meanwhile,
the complex initial position of complex worldline $x_0^\m(0)$
in (\ref{dec})
gives six parameters which are connected to coefficients
$a_0, \ a_1 \ a_2 \ $.  It can be decomposed as $\vec x_0 (0) = \vec c +
i\vec
d$, where $\vec c $ and $\vec d$ are real 3-vectors with respect to the
space
O(3)-rotation.  The real part $\vec c$ defines the initial position of
source,
and the imaginary part $\vec d$ defines the value and direction of angular
momentum.

\section{Real slice and the real retarded time}

In nonstationary case coefficients of function $F$ turn out to be
complex variable depending on the complex retarded-time parameter $\t$,
and function $\d _\tau x_0^\mu =\xi ^\mu $ takes also complex values.
 The real slice of space-time is constructed from
the ``left'' and ``right''complex structures. The ``left'' structure is
built of the left complex worldline $x_0$ and of the complex  parameter $Y$
generating the left null planes.
The  ``right'' complex
structure is built of the right complex worldline $\bar x_0$, parameter
$\Y$ and right null planes, spanned by vectors $e^2$ and $e^3$.  These
structures can be considered as functionally independent in $CM^4$, but they
have to be complex conjugate on the real slice of  space-time.
For a real point of space-time $x$ and for the corresponding real null
direction $e^3$ we define a real function
\be \rho (x) = x^\m e^3_\m (x).
 \label{rho} \ee
One can determine the values of $\rho$ at the points of the
left and right complex worldlines $x_0^\mu$ and $\bar x_0 ^\mu$ by L- and
R-projections \be \rho _L (x_0) =  x_0^\m e^3_\m (x)|_L, \qquad \rho _R(\bar
x_0) = \bar x_0^\m e^3_\m (x)|_R,\label{rhoR}\ee where the sign $|_L$ means
that the points $x$ and $x_0(\t)$ are synchronized by the left  null plane
(\ref{L}), and $ x
- x_0(\t_L) = \alpha e^1 + \beta e^3 . $

As a consequence of the
conditions $e^{1\mu} e^3_\mu=e^{3\mu} e^3_\mu =0$, we have
\be \rho _L(x_0) = x_0^\m e^3_\m (x)|_L = \rho (x).
\label{rho0}\ee
So far as the parameter
$\rho (x) $ is real, parameter $\rho _L (x_0)$ will be real too.
Similarly, $ \rho _R(\bar x_0) = {\bar x_0}^\m e^3_\m (x)|_R = \rho (x),$
and consequently, $ \rho _L(x_0) = \rho (x) = \rho _R (\bar x_0).$
The parameters  $\l_1,\quad \l_2$ can also be expressed
in terms of the coordinates $x^\mu$,
\be
\l_1 = x^\m e^1_\m , \qquad \l_2 = x^\m
(e^3_\m - \Y e^1_\m). \label{lambx}
\ee
It yields
\be \rho = \l_2 + \Y
\l_1 \ .  \label{rhotw} \ee
The values of twistor parameters $\l_1$ and $\l_2$ can also be defined by
L-projection for the points of the complex worldline,
$\l_1^0$ and $\l_2^0$:
 \be (\l_1-\l_1^0)|_L =
 0,\qquad (\l_2-\l_2^0)|_L =0 . \label{twL} \ee
The L-projection determines the  values of the left
retarded-time parameter $\tau _L= (t + i\sigma)|_L$.
The real function $\rho$ and the twistor variables
$ \l _1$ and $  \l _2 $
acquire an extra dependence on the retarded-time parameter $\tau _L$.
However, it should be
noted that the real and imaginary parts of $\tau _L$ are not independent
because of the constraint caused by L-projection. It means that {\it on the
real slice} functions $\rho , \quad \l_1$ and $\l_2$ and functions $\rho _0
,
 \quad \l_1^0, \quad \l_2^0$ can be considered as functions of the real
retarded-time parameter $t_0 =\Re e \  \tau _L$.

On the other hand, in $CM^4$ function $t_0 |_L$ is an analytic function of
twistor parameters $Z^a =\{Y,\quad \l _1, \quad \l _2 \}$ which satisfy the
relation\fn{It can be obtained by direct differentiation. See also
\cite{IvBur,BKP}} $Z^a ,_2 =Z^a ,_4=0$.  It has the consequence
\be t_0 ,_2 = t_0 ,_4 =0 \ .  \label{t02}\ee

Similar to the stationary case considered
in \cite{IvBur,BurMag}, one can use for function $F$
representation  in the form
\be F \equiv (\l_1 - \l_1^0) K_2 - (\l_2 -\l_2^0) K_1  ,
\label{Fnst} \ee
where the functions $K_1$ and $K_2$ will be depending
on the real retarded-time $t_0$ (or on the related real parameter $\rho_0$).
 It leads to the form (\ref{FK}) with the coefficients depending on the
retarded-time.

Let us assume that the relation $F(Y,t_0)=0$ is hold by the retarded-time
evolution $\d F / \d {t_0}|_L=0.$
It yields
\be \frac {\d F}{\d {t_0}}= K_1
\d_{t_0} \l_2^0 - K_2 \d_{t_0} \l_1^0  +
 (\l_1 -\l_1^0) \d_{t_0} K_2 - (\l_2
-\l_2^0) \d_{t_0} K_1 =0.
\label{dFdt0} \ee
As a consequence of (\ref{twL}),
by L-projection last two terms cancel and one obtains
\be ({\d F}/{\d {t_0}})|_L= (K_1 \d_{t_0} \l_2^0 - K_2 \d_{t_0}
\l_1^0)|_L =0,  \label{dFK12} \ee
that is provided by \be K_1 (t_0) =\d_{t_0}
 \l_2^0 ,\qquad K_2 (t_0) =\d_{t_0} \l_1^0.  \label{K12} \ee It seems that
the
extra dependence of function $F$ on the real retarded-time parameter $t_0$
contradicts to the Kerr theorem; however, the analytic dependence of $t_0$
on
 $Y,\quad \l _1, \quad \l _2$ is reconstructed by L-projection. As a result,
function F turns out to be analytic functions of twistor variables.
Meanwhile, all the {\it real} retarded-time derivatives are non-analytic and
have to involve the conjugate ``right'' complex structure.

As a consequence of the relation (\ref{PF}), one obtains
\be
P=\Y K_1 +K_2 , \label{PK12}
\ee
that yields for function $P$ the real expression
\be
P=\d \rho _L /\d t_0\ = \Re e \ \d _\t (x_0^\m (\t) e^3_\m |_L).\label{Pnst}
\ee
The coefficients $A,B,C$ of the resulting decomposition $F=AY^2 +BY +C$ will
be functions of the retarded-time parameter $t_0$ as well as the solutions
of (\ref{Y12}),(\ref{tr1}) defining $\tilde r$, $Y(x)$ and corresponding
PNC.

\fn{Note that the real function $\rho _L (t_0)$ plays
the role of a potential for $P$, Similar to some nonstationary solutions
presented in \cite{KraSte}}.

\section{Solution of the field equations. }

For simplicity, we shall assume that there is no electromagnetic field.
Similar to the Kinnersley case, we admit the existence of null radiation.
Therefore, all the components of Ricci tensor $R_{ab}$ have to be zero for
the exclusion $R_{33}$ that corresponds to an incoherent flow of the
light-like particles in $e^3$ direction.

The process of the solution of the field equations is  similar to the
treatment given in \cite{DKS}.  In particular, we have $ R_{24} =R_{22}
=R_{44}=0. $

If the electromagnetic field is zero, we have also
 $ R_{12} =R_{34}=0,$ that leads to the
equation $ h,_{44} + 2(Z + \cZ)h,_4 + 2Z \cZ h =0,$
which admits the solution
\be h= M(Z+ \cZ)/2  \ ,
\label{(5.5)}\ee
where $M$ is a real function,
obeying the condition $ M,_4=0 $.
\par
  The equation $ R_{23} = 0 , $  acquires the form
\be M,_2 - 3 Z^{-1} \cZ Y,_3 M  = 0.\label{7}\ee

  The last gravitational field equation
$R_{33} = -P_{33}$ takes the form
\be
\cD M =Z^{-1} \cZ ^{-1} P_{33}/2  , \label{last}\ee
where
\be
\cD = \d _3 - Z^{-1} Y,_3 \d _1 - \cZ ^{-1} \Y ,_3 \d _2 . \label{cD}\ee
The term $P_{33}$ is the contribution to energy-momentum tensor
corresponding to the null radiation.
\par
To integrate (\ref{7}) we have to
use the relation \fn{This relation was proved in \cite{DKS} for stationary
case.
In nonstationary case function $P$ has an extra dependence on $t_0$;
however, because of (\ref{t02}) this relation is valid.}
\be (\log P),_2 =-Z^{-1}\cZ Y,_3
\label{PY3},  \ee
which allows us to represent (\ref{7}) in the form
\be (\log MP^3),_2=0 \ee and to get the general solution
\be M= m/P^3, \label{11}\ee
where
\be m,_4 = m,_2 =0.  \label{(5.12)}\ee
Since $m$ has to be real, it can only be a function of $t_0$.

  Action of the operator $\cD$ on the variables
$Y, \bar Y $ and $ \rho$ is as follows: \be \cD Y = \cD \bar Y = 0,\quad
\cD \rho =1. \label{(5.16)}\ee
From the last relation and (\ref{Pnst}) we have
\be
\cD \rho =(\d \rho _L / \d t_0) \cD t_0 =P \cD t_0 =1 ,
\ee
which yields
\be \cD t_0 = P^{-1} .  \ee
Since $ M$ is a function of $Y, \bar Y$ and $ t_0$,  equation
(\ref{last}) takes the form
\be \partial _{t_0} M = P  Z^{-1} \cZ ^{-1}
P_{33}/2 . \label{5.17}\ee
This equation is the
definition of the unique component $P_{33}$ of the lightlike radiation which
propagates along twisting PNC.
Substituting (\ref{11}) one obtains
\be P_{33} = Z \cZ
[-6m(\partial _{t_0} P) + 2 P(\partial _{t_0} m)]/P^3.
\label{P33}\ee
The first term describes a radiation caused by an acceleration of source
while the second term corresponds to a radiation with a loss of mass
that corresponds to the Vaidya "shining star" solution
\cite{KraSte,VaiPat,FroKhl}.

The resulting metric has the form
$ g_{\m\n} = \h_{\m\n} + (m/P^3)(Z +\bar Z)
   e^3_{\m} e^3_{\n} \ $.
One can normalize $e^3$ by introducing
$l = e^3/P$,  and
metric takes the simple form
\be g_{\m\n} =\h_{\m\n} - m({\tilde r}^{-1} +
\bar{\tilde r}^{-1}) l_{\m} l_{\n}, \label{(5.21)}\ee where
$\tilde r =  PZ^{-1} = - dF/dY .$

The structure of this solution and the form of metric are similar to the
Kinnersley solution \cite{Kin} and correspond to its
modification for the case of complex worldline.

Let us summarize the obtained solution. All the parameters are determined
by a given complex worldline $x_0(\tau)$ and have the
current values depending on the retarded-time parameter $t_0$ which is
determined by L-projection
 \be (\l_1-\l_1^0)|_L = 0,\qquad (\l_2-\l_2^0)|_L =0 . \label{s1} \ee
The congruence
\be
 e^3 = du+ \Y d \z  + Y d \Z - Y \Y d v \
\label{s2}
\ee
is expressed in the null Cartesian coordinates $u,v,\z,\Z$ and is
determined for a fixed retarded time $t_0$ by function
\be Y (x,t_0) = [- B + (B^2 -4AC)^{1/2} ]/2A \label{s3} \ee
which is a solution of the equation $F=AY^2+BY +C=0$.
 The current parameters $A,B,C$ are defined by this, quadratic in $Y$,
decomposition of function $F$ given by
\be F \equiv (\l_1 - \l_1^0) K_2 - (\l_2 -\l_2^0) K_1  , \label{s4} \ee
where \be K_1 (t_0) =\d_{t_0} \l_2^0 ,\qquad K_2 (t_0) =\d_{t_0} \l_1^0.
 \label{s5} \ee
The current function $\tilde r$ and $P$ are given by \be\tilde
r = - 2AY-B , \quad P=\Y K_1 +K_2 . \label{s6} \ee

If the worldline is real, $ \Im m \ x_0 = 0, $ the equation
(\ref{11}) takes the form $ M (Y,\bar Y, t_0) = m/P^3,$ where
$P=\dot x_0^\m e^3_\m$ so that $\dot x_0^\m l_\m=1$.
We obtain $\tau _L= \tau _R$ that yields exactly the Kinnersley
real retarded-time construction with metric
$ g_{\m\n} = \h_{\m\n} +
2(m/r) (\sigma_\m /r) (\sigma_\n /r). $ The relation with our
notations is   $ l^\m = \sigma^\m/r, $ where $ \sigma^\m = x^\m -
x_0^\m,\qquad r = PZ^{-1}.$  The Kinnersley retarded-time parameter $ u =
\tau/ \sqrt {(\dot x_0)^2},$ and the Kinnersley parameter $\lambda^\m (u) =
\dot x_0^\m (\tau)/ \sqrt{(\dot x_0)^2}.$

\section{Conclusion.}
The solutions considered represent a natural generalization of
the Kinnersley class to the rotating case.
The Kerr-Schild approach provides the exact retarded expressions for metric,
coordinate system, PNC, and position of singularity in terms of the
null Cartesian coordinates by arbitrary boost and acceleration of rotating
source with arbitrary orientations of angular momentum.

 The solutions obtained can find an application for modelling  the
radiation from the rotating astrophysical object by arbitrary relativistic
boosts and accelerations and can also represent an interest for
investigation
of the role of gravitational field by particle scattering in
ultrarelativistic regimes \cite{BurMag}, as well as for  modelling
the Kerr spinning particle \cite{IvBur,Bur1,BurSup,KN},
generating excitations of the string-like Kerr source \cite{BurSen}.
On the other hand, the solutions considered are direct
generalizations of the rotating black hole solutions, and have the black
hole
horizons by the standard conditions $m^2>e^2+a^2$.  Since the black hole
solutions in the retarded-time scheme are based on outgoing principal null
congruence, the black hole interior has to be considered in this case as the
`past'. The problem of radiation from black holes by acceleration demands a
better understanding of the corresponding geometry and has to be considered
elsewhere.

Some known generalizations of the Kerr solution, such as
Kerr-Newman solution, Kerr-Sen solution to low energy string
theory \cite{BurSen}, solution to broken $N=2$ supergravity
\cite{BurSup} and the regular rotating BH-solutions \cite{BEHM},
retain the form and (geodesic and shear-free) properties of the Kerr
congruence. It means that the Kerr theorem can be used in these cases,
and these solutions can also be generalized to the nonstationary cases.

{\bf Acknowledgments.}
We are very thankful to G. Alekseev and to M. Demianski
for detailed discussions.

\end{document}